\documentstyle[graphicx,12pt]{article}
\begin{document}

\small
\hoffset=-1truecm
\voffset=-2truecm
\title{\bf The Casimir effect for parallel plates at finite temperature
in the presence of one fractal extra compactified dimension}
\author{Hongbo Cheng\footnote {E-mail address:
hbcheng@sh163.net}\\
Department of Physics, East China University of Science and
Technology,\\ Shanghai 200237, China\\
The Shanghai Key Laboratory of Astrophysics,\\ Shanghai 200234,
China}

\date{}
\maketitle

\begin{abstract}
We discuss the Casimir effect for massless scalar fields subject
to the Dirichlet boundary conditions on the parallel plates at
finite temperature in the presence of one fractal extra
compactified dimension. We obtain the Casimir energy density with
the help of the regularization of multiple zeta function with one
arbitrary exponent and further the renormalized Casimir energy
density involving the thermal corrections. It is found that when
the temperature is sufficiently high, the sign of the Casimir
energy remains negative no matter how great the scale dimension
$\delta$ is within its allowed region. We derive and calculate the
Casimir force between the parallel plates affected by the fractal
additional compactified dimension and surrounding temperature. The
stronger thermal influence leads the force to be stronger. The
nature of the Casimir force keeps attractive.
\end{abstract}
\vspace{4cm} \hspace{1cm} PACS number(s): 03.70.+k, 03.65.Ge

\newpage

\noindent \textbf{I.\hspace{0.4cm}Introduction}

The model of higher-dimensional spacetime as a powerful ingredient
is employed to unify the interactions in nature. In the
Kaluza-Klein theory, one extra dimension in our Universe was
introduced to be compactified in order to unify gravity and
classical electrodynamics [1, 2]. The characteristic size of the
additional dimension is of the order of Planck length. It may be
better to describe the background whose scale is on the Planck
order with the help of fractal geometry involving some non-integer
dimensions in the process of the investigation of quantum gravity
[3]. More attentions of the physical community are attracted to
this topic. It is found that the spectral dimension of the
spacetime where Quantum Einstein Gravity lives in is equal to 2
microscopically while to 4 on macroscopic scales [4]. It is also
discovered that the background with a quantum symmetry has a
scale-dependent fractal dimension at short scales to describe a
phenomenon appeared in the quantum gravity [5]. A kind of field
theory which is Lorentz invariant, power-counting renormalizable,
ultraviolet finite and causal is proposed to explore a consistent
theory of quantum gravity needing the fractal geometry [6, 7].
Within the frame of modern Kaluza-Klein issue it seems to be
reasonable to describe the extra compactified space in virtue of
fractal geometry because the quantum fluctuations is extremely
tiny close to the Planck scale. I. Smolyaninov originated a
fractal extra compactified dimension to modify the Kaluza-Klein
model [8].

More than 60 years ago Casimir put forward an effect of boundaries
[9]. The so-called Casimir effect is essentially a direct
consequence of quantum field theory due to a change in the
spectrum of vacuum oscillations when the quantization volume is
bounded, or some background fields are inserted. More than 40
years ago Boyer found that the Casimir force for a conducting
spherical shell is repulsive, meaning that the circumstance will
determine the nature of the force [10]. Afterwards, a lot of
effort has been contributed to the related topics and more and
more results and methods have been put forward [11-18]. The
precision of the measurements has been greatly improved
experimentally [19-21]. The Casimir effect has something to do
with various factors, so the sign of the Casimir energy and the
nature of the Casimir force can become principles to be applied in
many subjects. The Casimir effect can be used to explore
high-dimensional spacetimes. We can study the Casimir effect for
the simple device such as parallel plates in the spacetimes with
extra compactified dimensions to show that the extra-dimension
influence was manifest and distinct, then we open a window to
probe the Kaluza-Klein model [22-37]. Recently more attention of
the physical community is also paid to the Casimir effect for
parallel plates or piston in the braneworld, such as
Randall-Sundrum models, etc. [38-51]. During the investigation
above we can estimate the properties of the warped world which may
be utilized to resolve the hierachy problem. The Casimir effect
has also been discussed in the context of string theory [52-54].
In addition the Casimir effect for fermionic field within the
parallel plates with various types of boundary conditions was
evaluated [55, 56].

It is fundamental to research on the Casimir effect for parallel
plates in the presence of fractal extra compactified dimensions.
As mentioned above the size of additional space is extremely tiny
close to the Planck order, so the properties of the compactified
space should be described in virtue of fractal geometry. We have
studied the Casimir effect in the parallel-plate system in the
background involving one fractal extra compactified dimension
[57]. The number of the additional spatial dimension is fractal
instead of being exactly a positive integer, so the dimensionality
varies within a smaller region. Here the dimensionality of
additional space is denoted as $D=D_{T}+\delta$, where $D_{T}$ is
topological dimension and $\delta$ is the scale dimension. We show
that the negative sign of the renormalized Casimir energy which is
the difference between the regularized energy for two parallel
plates and the one with no plates narrows the region of the
fractal dimensionality of additional space, the scale dimension
$\delta\in(\frac{1}{2}, 1)$ instead of $\delta\in[0, 1)$. In
addition, it is also found that the larger scale dimension will
lead to the greater revision on the original Casimir force. The
Casimir force between two parallel plates can show whether the
dimensionality of additional space is integer or fraction because
the shapes of two kinds of Casimir force are not exactly
identical.

The quantum field theory at finite temperature shares a lot of
effects. The thermal influence on the Casimir effect can not be
neglected, and its influence certainly modifies the effect. The
Casimir energy for a rectangular cavity in a background with
nonzero temperature was evaluated, and the temperature controls
the energy sign [58]. The Casimir effect for parallel plates,
including thermal corrections in the world with additional
compactified dimensions, was discussed, and the magnitude of
Casimir force as well as the sign of Casimir energy relates to the
temperature [59-63]. The Casimir effect for a scalar field within
two parallel plates under thermal influence in the bulk region of
Randall-Sundrum models was studied [64]. We have researched on the
Casimir effect for parallel plates involving massless Majorana
fermions obeying the bag boundary conditions at finite temperature
[65]. The Casimir force and Casimir free energy instead of Casimir
energy for massless Majorana fermions with thermal modifications
in a magnetic field is also investigated [66].

Here we plan to discuss the thermal corrections to the Casimir
effect for parallel plates in the spacetime with a fractal extra
compactified dimension in detail to generalize the results of Ref.
[57]. We wonder whether the thermal influence enlarges or narrows
the range belonging to the scale dimension of additional
compactified space and how the influence revises the description
of the Casimir effect for parallel plates under this kind of
environment. At first we derive the frequency of massless scalar
fields referring to Dirichlet boundary conditions at plates
containing thermal influence by means of the
fractal-extra-dimension Kaluza-Klein model put forward by
Smolyaninov [8] and finite-temperature field theory. We regulariza
the total vacuum energy density to obtain the Casimir energy
density by means of the regularization technique of multiple zeta
function with arbitrary exponents. During this investigation we
obtain the finite part of this kind of zeta function according to
the procedure in Ref. [13, 14]. Further we obtain the renormalized
Casimir energy density which is the difference between the
regularized energy density for two parallel plates and the one
with no plates at finite temperature on purpose to subtract the
divergent part. The Casimir force between the parallel plates can
also be gained from the renormalized Casimir energy density. Our
discussions for thermal influence on this kind of Casimir effect
are given at the end of this paper.

\vspace{0.8cm} \noindent \textbf{II.\hspace{0.4cm}The Casimir
energy for parallel plates at finite temperature in the spacetime
with a fractal extra compactified dimension}

Within the Kaluza-Klein issue with a fractal extra compactified
dimension [8], the scalar field is periodic in the fifth
coordinate $x^{5}$, leading to the appearance of an infinite tower
of solutions with a quantized $x^{5}$-component of the momentum
like $q_{n}=\frac{2\pi n}{L}$. The scale dimension is defined as,

\begin{eqnarray}
\delta=D-D_{T}\nonumber\\\equiv\frac{d(\ln
L)}{d(\ln\frac{l}{\lambda})}\hspace{0.5cm}
\end{eqnarray}

\noindent where $L$ is the main fractal variable denoting a length
of a fractal curve, an area of a fractal surface, etc.. The
coefficients $l$ and $\lambda$ are measurement scale. Here $D$ is
the fractal dimensionality. $D_{T}$ is the topological dimension
and $D_{T}=1$ for a curve, $D_{T}=2$ for a surface. The spectrum
of momentum is,

\begin{equation}
q_{n}=2\pi(\frac{n}{L_{0}l^{\delta}})^{\frac{1}{1+\delta}}
\end{equation}

\noindent where $n$ is a nonnegative integer. $L_{0}$ is the
length measured when $\lambda=l$. This tower of solutions will
recover to be the tower in the regular -five-dimensional
Kaluza-Klein theory if we choose $\delta=0$.

In finite-temperature field theories the imaginary time formalism
can be employed to describe the scalar fields in thermal
equilibrium. We introduce a partition function for a system,

\begin{equation}
Z=N\int_{periodic}\prod_{k}D\phi_{k}\exp[\int_{0}^{\beta}d\tau
\int d^{3}x \cal{L}(\phi_{k}, \partial_{E}\phi_{k})]
\end{equation}

\noindent where $\cal{L}$ is the Lagrangian density for the system
under consideration. $N$ is a constant and the "periodic" means
$\phi_{k}(0, \mathbf{x})=\phi_{k}(\tau=\beta, \mathbf{x})$, $k=0,
1, 2, \cdot\cdot\cdot$, and $\beta=\frac{1}{T}$ is the inverse of
the temperature and $\tau=it$. The scalar fields $\phi_{k}$
satisfy the Klein-Gordon equations
$(\partial_{\mu}\partial^{\mu}-\frac{k^{2}}{L^{2}})\phi_{k}(x)=0$.
The fields confined between the two parallel plates obey the
Dirichlet boundary conditions $\phi_{k}(x)|_{\partial\Omega}=0$,
and $\partial\Omega$ stands for positions of the plates. According
to the solutions to the Klein-Gordon equation and the boundary
conditions, the generalized zeta function can be written as,

\begin{eqnarray}
\zeta(s; -\partial_{E})=Tr(-\partial_{E})^{-s}\nonumber\hspace{6.5cm}\\
=\int\frac{d^{2}\kappa}{(2\pi)^{2}}\sum_{n=1}^{\infty}
\sum_{n_{1}=0}^{\infty}\sum_{l=-\infty}^{\infty}
[\kappa^{2}+\frac{n^{2}\pi^{2}}{R^{2}}+\frac{(2\pi)^{2}}{(L_{0}l^{\delta})^{\frac{2}{1+\delta}}}
n_{1}^{\frac{2}{1+\delta}}+(\frac{2l\pi}{\beta})^{2}]^{-s}
\end{eqnarray}

\noindent where

\begin{equation}
\partial_{E}=\frac{\partial^{2}}{\partial\tau^{2}}+\nabla^{2}
\end{equation}

\noindent and $\kappa^{2}=k_{1}^{2}+k_{2}^{2}$ denote the
transverse components of the momentum. $R$ is the separation of
plates. The total energy density of the system with thermal
corrections is,

\begin{eqnarray}
\varepsilon=-\frac{\partial}{\partial\beta}(\frac{\partial\zeta(s;
-\partial_{E})}{\partial s}|_{s=0})\nonumber\hspace{6cm}\\
=-\frac{1}{2\pi}\Gamma(-1)\frac{\partial}{\partial\beta}E_{2}(-1;
\frac{\pi^{2}}{R^{2}}, \frac{4\pi^{2}}{\beta^{2}})\nonumber\hspace{5cm}\\
-\frac{1}{2\pi}\Gamma(-1)\frac{\partial}{\partial\beta} M_{3}(-1;
\frac{\pi^{2}}{R^{2}}, \frac{4\pi^{2}}{\beta^{2}},
\frac{(2\pi)^{2}}{(L_{0}l^{\delta})^{\frac{2}{1+\delta}}}; 2, 2,
\frac{2}{1+\delta})
\end{eqnarray}

\noindent in terms of Epstein zeta function $E_{2}(s; a_{1},
a_{2})=\sum_{n_{1},n_{2}=1}^{\infty}(a_{1}n_{1}^{2}+a_{2}n_{2}^{2})^{-s}$
and the multiple zeta function with arbitrary exponents
$M_{3}(s;a_{1},a_{2},a_{3};\alpha_{1},\alpha_{2},\alpha_{3})$
defined as,

\begin{equation}
M_{3}(s;a_{1},a_{2},a_{3};\alpha_{1},\alpha_{2},\alpha_{3})
=\sum_{n_{1},n_{2},n_{3}=1}^{\infty}(a_{1}n_{1}^{\alpha_{1}}
+a_{2}n_{2}^{\alpha_{2}}+a_{3}n_{3}^{\alpha_{3}})^{-s}
\end{equation}

\noindent Here we make use of the standard method to regularize
the multiple zeta function with one arbitrary exponent
$M_{3}(s;a_{1},a_{2},a_{3};2,2,\alpha_{3})$ which will be used in
this work as follow,

\begin{eqnarray}
M_{3}(s;a_{1},a_{2},a_{3};2,2,\alpha_{3})\nonumber\hspace{9cm}\\
=\sum_{n_{1},n_{2},n_{3}=1}^{\infty}(a_{1}n_{1}^{2}
+a_{2}n_{2}^{2}+a_{3}n_{3}^{\alpha_{3}})^{-s}\nonumber\hspace{7.4cm}\\
=-\frac{1}{2}M_{2}(s;a_{2},a_{3};2,\alpha_{3})
+\frac{1}{2}\sqrt{\frac{\pi}{a_{1}}}\frac{\Gamma(s-\frac{1}{2})}{\Gamma(s)}
M_{2}(s-\frac{1}{2};a_{2},a_{3};2,\alpha_{3})\nonumber\hspace{2.5cm}\\
+\frac{2}{\Gamma(s)}\sqrt{\frac{\pi}{a_{1}}}\sum_{n_{1},n_{2},n_{3}=1}^{\infty}
(\frac{n_{1}\pi}{\sqrt{a_{1}(a_{2}n_{2}^{2}+a_{3}n_{3}^{\alpha_{3}})}})^{s-\frac{1}{2}}
K_{-s+\frac{1}{2}}(2\pi
n_{1}\sqrt{\frac{a_{2}n_{2}^{2}+a_{3}n_{3}^{\alpha_{3}}}{a_{1}}})
\end{eqnarray}

\noindent which contains the tiny contribution denoted as the
modified Bessel function term. The multiple zeta function
$M_{2}(s;a_{1},a_{2};2,\alpha_{2})$ has also been regularized [57]
and is expressed as,

\begin{eqnarray}
M_{2}(s; a_{1}, a_{2}; 2, \alpha_{2})\nonumber\hspace{7cm}\\
=\sum_{n_{1},n_{2}=1}^{\infty}(a_{1}n_{1}^{2}
+a_{2}n_{2}^{\alpha_{2}})^{-s}\nonumber\hspace{6cm}\\
=-\frac{1}{2}a_{2}^{-s}\zeta(\alpha_{2}s)+\frac{1}{2}
a_{2}^{-s}\sqrt{\frac{\pi
a_{2}}{a_{1}}}\frac{\Gamma(s-\frac{1}{2})}{\Gamma(s)}
\zeta(\alpha_{2}(s-\frac{1}{2}))\nonumber\hspace{1.7cm}\\
+2\pi^{s}a_{1}^{-\frac{s}{2}-\frac{1}{4}}a_{2}^{-\frac{s}{2}+\frac{1}{4}}
\sum_{n_{1},n_{2}=1}^{\infty}(\frac{n_{1}}{n_{2}^{\frac{\alpha_{2}}{2}}})^{s-\frac{1}{2}}
K_{-(s-\frac{1}{2})}(2\pi\sqrt{\frac{a_{2}}{a_{1}}}n_{1}n_{2}^{\frac{\alpha_{2}}{2}})
\end{eqnarray}

\noindent where $K_{\nu}(z)$ is the modified Bessel function of
the second kind and drops exponentially with $z$. According to the
regularization of the multiple zeta functions with one arbitrary
exponent in Eq.(8) and (9), we rewrite the total energy density of
the two parallel plates at finite temperature as,

\begin{eqnarray}
\varepsilon=-\frac{\pi^{2}}{720}\frac{1}{R^{3}}
+\frac{1}{\sqrt{2}}\frac{1}{(\beta R)^{\frac{3}{2}}}
\sum_{n_{1},n_{2}=1}^{\infty}(\frac{n_{2}}{n_{1}})^{\frac{3}{2}}
K_{\frac{3}{2}}(\pi\frac{\beta}{R}n_{1}n_{2})\nonumber\hspace{3.5cm}\\
+\frac{\pi}{\sqrt{2}}\frac{1}{\beta^{\frac{1}{2}}R^{\frac{5}{2}}}
\sum_{n_{1},n_{2}=1}^{\infty}\frac{n_{2}^{\frac{5}{2}}}{n_{1}^{\frac{1}{2}}}
[K_{\frac{1}{2}}(\pi\frac{\beta}{R}n_{1}n_{2})
+K_{\frac{5}{2}}(\pi\frac{\beta}{R}n_{1}n_{2})]\nonumber\hspace{2.5cm}\\
+\frac{\pi^{\frac{3}{2}}}{2}\Gamma(-\frac{3}{2})\zeta(-\frac{3}{1+\delta})
\frac{1}{(L_{0}l^{\delta})^{\frac{3}{1+\delta}}}\nonumber\hspace{6cm}\\
-\frac{1}{\beta^{\frac{3}{2}}(L_{0}l^{\delta})^{\frac{3}{2(1+\delta)}}}
\sum_{n_{1},n_{2}=1}^{\infty}(\frac{n_{2}^{\frac{1}{1+\delta}}}{n_{1}})^{\frac{3}{2}}
K_{\frac{3}{2}}(2\pi\frac{\beta}{(L_{0}l^{\delta})^{\frac{1}{1+\delta}}}n_{1}n_{2}^{\frac{1}{1+\delta}})\nonumber\hspace{2cm}\\
-\frac{2\pi}{\beta^{\frac{1}{2}}(L_{0}l^{\delta})^{\frac{5}{2(1+\delta)}}}
\sum_{n_{1},n_{2}=1}^{\infty}\frac{n_{2}^{\frac{5}{2(1+\delta)}}}{n_{1}^{\frac{1}{2}}}
[K_{\frac{1}{2}}(2\pi\frac{\beta}{(L_{0}l^{\delta})^{\frac{1}{1+\delta}}}n_{1}n_{2}^{\frac{1}{1+\delta}})\nonumber\hspace{2cm}\\
+K_{\frac{5}{2}}(2\pi\frac{\beta}{(L_{0}l^{\delta})^{\frac{1}{1+\delta}}}n_{1}n_{2}^{\frac{1}{1+\delta}})]\nonumber\hspace{5cm}\\
-4\pi^{-\frac{3}{2}}\Gamma(-\frac{3}{2})R\frac{\partial}{\partial\beta}
M_{2}(-\frac{3}{2};\frac{4\pi^{2}}{\beta^{2}},\frac{(2\pi)^{2}}{(L_{0}l^{\delta})^{\frac{2}{1+\delta}}};
2,\frac{2}{1+\delta})\nonumber\hspace{2.5cm}\\
+\frac{6\pi^{\frac{1}{2}}}{R^{\frac{1}{2}}\beta^{3}}
\sum_{n_{1},n_{2},n_{3}=1}^{\infty}\frac{n_{2}^{2}}{n_{1}^{\frac{3}{2}}}
(\frac{4\pi^{2}}{\beta^{2}}n_{2}^{2}+\frac{(2\pi)^{2}}{(L_{0}l^{\delta})^{\frac{2}{1+\delta}}}
n_{3}^{\frac{2}{1+\delta}})^{-\frac{1}{4}}\nonumber\hspace{3cm}\\
\times
K_{\frac{3}{2}}(2n_{1}R\sqrt{\frac{4\pi^{2}}{\beta^{2}}n_{2}^{2}+\frac{(2\pi)^{2}}{(L_{0}l^{\delta})^{\frac{2}{1+\delta}}}
n_{3}^{\frac{2}{1+\delta}}})\nonumber\hspace{4cm}\\
-4\pi^{\frac{1}{2}}\frac{R^{\frac{1}{2}}}{\beta^{3}}
\sum_{n_{1},n_{2},n_{3}=1}^{\infty}\frac{n_{2}^{2}}{n_{1}^{\frac{1}{2}}}
(\frac{4\pi^{2}}{\beta^{2}}n_{2}^{2}+\frac{(2\pi)^{2}}{(L_{0}l^{\delta})^{\frac{2}{1+\delta}}}
n_{3}^{\frac{2}{1+\delta}})^{\frac{1}{4}}\nonumber\hspace{3cm}\\
\times[K_{\frac{1}{2}}(\frac{4\pi^{2}}{\beta^{2}}n_{2}^{2}+\frac{(2\pi)^{2}}{(L_{0}l^{\delta})^{\frac{2}{1+\delta}}}
n_{3}^{\frac{2}{1+\delta}})+K_{\frac{5}{2}}(\frac{4\pi^{2}}{\beta^{2}}n_{2}^{2}+\frac{(2\pi)^{2}}{(L_{0}l^{\delta})^{\frac{2}{1+\delta}}}
n_{3}^{\frac{2}{1+\delta}})]
\end{eqnarray}

\noindent In the absence of plates the vacuum energy density at
finite temperature is,

\begin{eqnarray}
\varepsilon_{0}=-\frac{\partial}{\partial\beta}(\frac{\partial}{\partial
s}\{\int\frac{d^{3}k}{(2\pi)^{3}}\sum_{n=0}^{\infty}\sum_{l=-\infty}^{\infty}
[k^{2}+\frac{(2\pi)^{2}}{(L_{0}l^{\delta})^{\frac{2}{1+\delta}}}n^{\frac{2}{1+\delta}}
+(\frac{2l\pi}{\beta})^{2}]^{-s}\})|_{s=0}\nonumber\\
=\frac{\pi^{2}}{15}\frac{1}{\beta^{4}}-\frac{1}{4\pi^{\frac{3}{2}}}
\Gamma(-\frac{3}{2})\frac{\partial}{\partial\beta}
M_{2}(-\frac{3}{2};\frac{4\pi^{2}}{\beta^{2}},\frac{(2\pi)^{2}}{(L_{0}l^{\delta})^{\frac{2}{1+\delta}}}
;2,\frac{2}{1+\delta})\hspace{2cm}
\end{eqnarray}

\noindent We can subtract the part of energy density for no plates
to renormalize the two-parallel-plate system energy density
denoted in Eq. (10). The terms like $-\frac{1}{4\pi^{\frac{3}{2}}}
\Gamma(-\frac{3}{2})\frac{\partial}{\partial\beta}
M_{2}(-\frac{3}{2};\frac{4\pi^{2}}{\beta^{2}},\frac{(2\pi)^{2}}{(L_{0}l^{\delta})^{\frac{2}{1+\delta}}}
;2,\frac{2}{1+\delta})$ in Eq. (10) and Eq. (11) respectively are
just compensated. We regularize the difference between the two
energy densities with or without plates respectively to obtain the
renormalized Casimir energy density,

\begin{eqnarray}
\varepsilon_{C}^{ren}=\varepsilon-R\varepsilon_{0}\nonumber\hspace{7.5cm}\\
=-\frac{\pi^{2}}{720}\frac{1}{R^{3}}
+\frac{\pi^{\frac{3}{2}}}{2}\Gamma(-\frac{3}{2})\zeta(-\frac{3}{1+\delta})
\frac{1}{(L_{0}l^{\delta})^{\frac{3}{1+\delta}}}
-\frac{\pi^{2}}{15}\frac{R}{\beta^{4}}\nonumber\hspace{1.5cm}\\
+\frac{1}{\sqrt{2}}\frac{1}{(\beta R)^{\frac{3}{2}}}
\sum_{n_{1},n_{2}=1}^{\infty}(\frac{n_{2}}{n_{1}})^{\frac{3}{2}}
K_{\frac{3}{2}}(\pi\frac{\beta}{R}n_{1}n_{2})\nonumber\hspace{2cm}\\
+\frac{\pi}{\sqrt{2}}\frac{1}{\beta^{\frac{1}{2}}R^{\frac{5}{2}}}
\sum_{n_{1},n_{2}=1}^{\infty}\frac{n_{2}^{\frac{5}{2}}}{n_{1}^{\frac{1}{2}}}
[K_{\frac{1}{2}}(\pi\frac{\beta}{R}n_{1}n_{2})
+K_{\frac{5}{2}}(\pi\frac{\beta}{R}n_{1}n_{2})]\nonumber\hspace{0.5cm}\\
-\frac{1}{\beta^{\frac{3}{2}}(L_{0}l^{\delta})^{\frac{3}{2(1+\delta)}}}
\sum_{n_{1},n_{2}=1}^{\infty}(\frac{n_{2}^{\frac{1}{1+\delta}}}{n_{1}})^{\frac{3}{2}}
K_{\frac{3}{2}}(2\pi\frac{\beta}{(L_{0}l^{\delta})^{\frac{1}{1+\delta}}}n_{1}n_{2}^{\frac{1}{1+\delta}})\nonumber\\
-\frac{2\pi}{\beta^{\frac{1}{2}}(L_{0}l^{\delta})^{\frac{5}{2(1+\delta)}}}
\sum_{n_{1},n_{2}=1}^{\infty}\frac{n_{2}^{\frac{5}{2(1+\delta)}}}{n_{1}^{\frac{1}{2}}}
[K_{\frac{1}{2}}(2\pi\frac{\beta}{(L_{0}l^{\delta})^{\frac{1}{1+\delta}}}n_{1}n_{2}^{\frac{1}{1+\delta}})\nonumber\\
+K_{\frac{5}{2}}(2\pi\frac{\beta}{(L_{0}l^{\delta})^{\frac{1}{1+\delta}}}n_{1}n_{2}^{\frac{1}{1+\delta}})]\nonumber\hspace{3cm}\\
+\frac{6\pi^{\frac{1}{2}}}{R^{\frac{1}{2}}\beta^{3}}
\sum_{n_{1},n_{2},n_{3}=1}^{\infty}\frac{n_{2}^{2}}{n_{1}^{\frac{3}{2}}}
(\frac{4\pi^{2}}{\beta^{2}}n_{2}^{2}+\frac{(2\pi)^{2}}{(L_{0}l^{\delta})^{\frac{2}{1+\delta}}}
n_{3}^{\frac{2}{1+\delta}})^{-\frac{1}{4}}\nonumber\hspace{1cm}\\
\times
K_{\frac{3}{2}}(2n_{1}R\sqrt{\frac{4\pi^{2}}{\beta^{2}}n_{2}^{2}+\frac{(2\pi)^{2}}{(L_{0}l^{\delta})^{\frac{2}{1+\delta}}}
n_{3}^{\frac{2}{1+\delta}}})\nonumber\hspace{1cm}\\
-4\pi^{\frac{1}{2}}\frac{R^{\frac{1}{2}}}{\beta^{3}}
\sum_{n_{1},n_{2},n_{3}=1}^{\infty}\frac{n_{2}^{2}}{n_{1}^{\frac{1}{2}}}
(\frac{4\pi^{2}}{\beta^{2}}n_{2}^{2}+\frac{(2\pi)^{2}}{(L_{0}l^{\delta})^{\frac{2}{1+\delta}}}
n_{3}^{\frac{2}{1+\delta}})^{\frac{1}{4}}\nonumber\hspace{1cm}\\
\times[K_{\frac{1}{2}}(2n_{1}R\sqrt{\frac{4\pi^{2}}{\beta^{2}}n_{2}^{2}+\frac{(2\pi)^{2}}{(L_{0}l^{\delta})^{\frac{2}{1+\delta}}}
n_{3}^{\frac{2}{1+\delta}}})\nonumber\hspace{1cm}\\
+K_{\frac{5}{2}}(2n_{1}R\sqrt{\frac{4\pi^{2}}{\beta^{2}}n_{2}^{2}+\frac{(2\pi)^{2}}{(L_{0}l^{\delta})^{\frac{2}{1+\delta}}}
n_{3}^{\frac{2}{1+\delta}}})]
\end{eqnarray}

\noindent If one of the plates is moved to the remote place, the
renormalized Casimir energy density reduces to be,

\begin{eqnarray}
\lim_{R\longrightarrow\infty}\varepsilon_{C}^{ren}\nonumber\hspace{9cm}\\
=-\frac{\zeta(3)}{\pi}\frac{1}{\beta^{3}}
-\frac{\pi^{\frac{\delta-2}{1+\delta}}}{2(1+\delta)}
\frac{\Gamma(\frac{3}{2(1+\delta)})}{\Gamma(\frac{3}{2})}
\Gamma(\frac{4+\delta}{2(1+\delta)})\zeta(\frac{4+\delta}{1+\delta})
\sin\frac{3\pi}{2(1+\delta)}\nonumber\\
-\frac{1}{\beta^{\frac{3}{2}}(L_{0}l^{\delta})^{\frac{3}{2(1+\delta)}}}
\sum_{n_{1},n_{2}=1}^{\infty}(\frac{n_{2}^{\frac{1}{1+\delta}}}{n_{1}})^{\frac{3}{2}}
K_{\frac{3}{2}}(2\pi\frac{\beta}{(L_{0}l^{\delta})^{\frac{1}{1+\delta}}}n_{1}n_{2}^{\frac{1}{1+\delta}})\nonumber\\
-\frac{2\pi}{\beta^{\frac{1}{2}}(L_{0}l^{\delta})^{\frac{5}{2(1+\delta)}}}
\sum_{n_{1},n_{2}=1}^{\infty}\frac{n_{2}^{\frac{5}{2(1+\delta)}}}{n_{1}^{\frac{1}{2}}}
[K_{\frac{1}{2}}(2\pi\frac{\beta}{(L_{0}l^{\delta})^{\frac{1}{1+\delta}}}n_{1}n_{2}^{\frac{1}{1+\delta}})
\end{eqnarray}

\noindent If we take $T=0$, our results in Eq.(13) will recover to
be those of our previous work [57] leading the constraint on the
scale dimension $\delta>\frac{1}{2}$ corresponding to the negative
nature of the Casimir energy for two parallel plates in the
presence of one fractal extra compactified dimension. It is
interesting that the sign of terms due to the temperature is
minus, so it is easier to keep the negative nature of the Casimir
energy. After numerical calculation we find that when we take
ratio
$\xi=\frac{(L_{0}l^{\delta})^{\frac{1}{1+\delta}}}{\beta}>0.54$,
the sign of the Casimir energy will remain negative no matter how
large the scale dimension is. We show the relation between $\xi$
and $\delta_{0}$ graphically in Fig. 1. For a definite temperature
or equivalently that the ratio $\xi$ has a definite value
belonging to $[0, 0.54]$, the sign of the Casimir energy for two
parallel plates in the world involving one fractal additional
compactified dimension keeps negative when the scale dimension for
extra space $\delta>\delta_{0}$. Fig. 1 demonstrates that the
parameter $\delta_{0}$ will decreases from $\frac{1}{2}$ to zero
when the ratio $\xi$ increases from zero to $0.54$. We can
emphasize that there is no limit on the scale dimension range like
$\delta\in(0,1)$ when the surrounding temperature is sufficiently
high.

\vspace{0.8cm} \noindent \textbf{III.\hspace{0.4cm}The Casimir
force between parallel plates at finite temperature in the
spacetime with a fractal extra compactified dimension}

It is important to continue studying the Casimir force within the
two-parallel-plate device involving the thermal influence when the
dimensionality of extra space governed by Kaluza-Klein theory is
not an integer. The Casimir force on the plates is given by the
derivative of the Casimir energy with respect to the plate
distance. Here the Casimir energy is thought as the renormalized
one. The Casimir force per unit area on the plates at finite
temperature in the presence of one fractal extra compactified
dimension can be written as,

\begin{eqnarray}
f_{C}=-\frac{\partial\varepsilon_{C}^{ren}}{\partial R}\nonumber\hspace{10cm}\\
=-\frac{\pi^{2}}{240}\frac{1}{R^{4}}+\frac{\pi^{2}}{15}\frac{1}{\beta^{4}}
+\frac{3}{2\sqrt{2}}\frac{1}{\beta^{\frac{3}{2}}R^{\frac{5}{2}}}
\sum_{n_{1},n_{2}=1}^{\infty}(\frac{n_{2}}{n_{1}})^{\frac{3}{2}}
K_{\frac{3}{2}}(\pi\frac{\beta}{R}n_{1}n_{2})\nonumber\hspace{2.4cm}\\
+\frac{\sqrt{2}\pi}{\beta^{\frac{1}{2}}R^{\frac{7}{2}}}
\sum_{n_{1},n_{2}=1}^{\infty}\frac{n_{2}^{\frac{5}{2}}}{n_{1}^{\frac{1}{2}}}
[K_{\frac{1}{2}}(\pi\frac{\beta}{R}n_{1}n_{2})
+K_{\frac{5}{2}}(\pi\frac{\beta}{R}n_{1}n_{2})]\nonumber\hspace{3.5cm}\\
-\frac{\pi^{2}}{2\sqrt{2}}\frac{\beta^{\frac{1}{2}}}{R^{\frac{9}{2}}}
\sum_{n_{1},n_{2}=1}^{\infty}n_{1}^{\frac{1}{2}}n_{2}^{\frac{7}{2}}
[K_{\frac{1}{2}}(\pi\frac{\beta}{R}n_{1}n_{2})
+2K_{\frac{3}{2}}(\pi\frac{\beta}{R}n_{1}n_{2})
+K_{\frac{7}{2}}(\pi\frac{\beta}{R}n_{1}n_{2})]\nonumber\\
+\frac{3\sqrt{\pi}}{R^{\frac{3}{2}}\beta^{3}}
\sum_{n_{1},n_{2},n_{3}=1}^{\infty}\frac{n_{2}^{2}}{n_{1}^{\frac{3}{2}}}
(\frac{4\pi^{2}}{\beta^{2}}n_{2}^{2}+\frac{(2\pi)^{2}}{(L_{0}l^{\delta})^{\frac{2}{1+\delta}}}
n_{3}^{\frac{2}{1+\delta}})^{-\frac{1}{4}}\nonumber\hspace{3.5cm}\\
\times
K_{\frac{3}{2}}(2n_{1}R\sqrt{\frac{4\pi^{2}}{\beta^{2}}n_{2}^{2}+\frac{(2\pi)^{2}}{(L_{0}l^{\delta})^{\frac{2}{1+\delta}}}
n_{3}^{\frac{2}{1+\delta}}})\nonumber\hspace{3cm}\\
+\frac{8\sqrt{\pi}}{R^{\frac{1}{2}}\beta^{3}}
\sum_{n_{1},n_{2},n_{3}=1}^{\infty}\frac{n_{2}^{2}}{n_{1}^{\frac{1}{2}}}
(\frac{4\pi^{2}}{\beta^{2}}n_{2}^{2}+\frac{(2\pi)^{2}}{(L_{0}l^{\delta})^{\frac{2}{1+\delta}}}
n_{3}^{\frac{2}{1+\delta}})^{\frac{1}{4}}\nonumber\hspace{4cm}\\
\times[K_{\frac{1}{2}}(2n_{1}R\sqrt{\frac{4\pi^{2}}{\beta^{2}}n_{2}^{2}+\frac{(2\pi)^{2}}{(L_{0}l^{\delta})^{\frac{2}{1+\delta}}}
n_{3}^{\frac{2}{1+\delta}}})\nonumber\hspace{3cm}\\
+K_{\frac{5}{2}}(2n_{1}R\sqrt{\frac{4\pi^{2}}{\beta^{2}}n_{2}^{2}+\frac{(2\pi)^{2}}{(L_{0}l^{\delta})^{\frac{2}{1+\delta}}}
n_{3}^{\frac{2}{1+\delta}}})]\nonumber\hspace{2cm}\\
-4\sqrt{\pi}\frac{R^{\frac{1}{2}}}{\beta^{3}}
\sum_{n_{1},n_{2},n_{3}=1}^{\infty}n_{1}^{\frac{1}{2}}n_{2}^{2}
(\frac{4\pi^{2}}{\beta^{2}}n_{2}^{2}+\frac{(2\pi)^{2}}{(L_{0}l^{\delta})^{\frac{2}{1+\delta}}}
n_{3}^{\frac{2}{1+\delta}})^{\frac{3}{4}}\nonumber\hspace{3.5cm}\\
\times[K_{\frac{1}{2}}(2n_{1}R\sqrt{\frac{4\pi^{2}}{\beta^{2}}n_{2}^{2}+\frac{(2\pi)^{2}}{(L_{0}l^{\delta})^{\frac{2}{1+\delta}}}
n_{3}^{\frac{2}{1+\delta}}})\nonumber\hspace{3cm}\\
+2K_{\frac{3}{2}}(2n_{1}R\sqrt{\frac{4\pi^{2}}{\beta^{2}}n_{2}^{2}+\frac{(2\pi)^{2}}{(L_{0}l^{\delta})^{\frac{2}{1+\delta}}}
n_{3}^{\frac{2}{1+\delta}}})\nonumber\hspace{2cm}\\
+K_{\frac{7}{2}}(2n_{1}R\sqrt{\frac{4\pi^{2}}{\beta^{2}}n_{2}^{2}+\frac{(2\pi)^{2}}{(L_{0}l^{\delta})^{\frac{2}{1+\delta}}}
n_{3}^{\frac{2}{1+\delta}}})]\hspace{2cm}
\end{eqnarray}

\noindent In Eq.(14) there is a term like
$\frac{\pi^{2}}{15}\frac{1}{\beta^{4}}$ which is the only one that
has nothing to do with plates separation $R$. If we employ the
piston model [30], this term will be compensated because the
$R$-independent forces acting on the plates have the same
magnitude and their sign are opposite each other. We rewrite the
Casimir force per unit area of plates as,

\begin{equation}
f_{C}=-\frac{\pi^{2}}{240}\frac{1}{R^{4}}+C(R, \beta,\delta)
\end{equation}

\noindent where the correction function $C(R, \beta,\delta)$ is
shown as,

\begin{eqnarray}
C(R, \beta,\delta)\nonumber\hspace{11cm}\\
=\frac{3}{2\sqrt{2}}\frac{1}{\beta^{\frac{3}{2}}R^{\frac{5}{2}}}
\sum_{n_{1},n_{2}=1}^{\infty}(\frac{n_{2}}{n_{1}})^{\frac{3}{2}}
K_{\frac{3}{2}}(\pi\frac{\beta}{R}n_{1}n_{2})\nonumber\hspace{5.8cm}\\
+\frac{\sqrt{2}\pi}{\beta^{\frac{1}{2}}R^{\frac{7}{2}}}
\sum_{n_{1},n_{2}=1}^{\infty}\frac{n_{2}^{\frac{5}{2}}}{n_{1}^{\frac{1}{2}}}
[K_{\frac{1}{2}}(\pi\frac{\beta}{R}n_{1}n_{2})
+K_{\frac{5}{2}}(\pi\frac{\beta}{R}n_{1}n_{2})]\nonumber\hspace{3.5cm}\\
-\frac{\pi^{2}}{2\sqrt{2}}\frac{\beta^{\frac{1}{2}}}{R^{\frac{9}{2}}}
\sum_{n_{1},n_{2}=1}^{\infty}n_{1}^{\frac{1}{2}}n_{2}^{\frac{7}{2}}
[K_{\frac{1}{2}}(\pi\frac{\beta}{R}n_{1}n_{2})
+2K_{\frac{3}{2}}(\pi\frac{\beta}{R}n_{1}n_{2})
+K_{\frac{7}{2}}(\pi\frac{\beta}{R}n_{1}n_{2})]\nonumber\\
+\frac{3\sqrt{\pi}}{R^{\frac{3}{2}}\beta^{3}}
\sum_{n_{1},n_{2},n_{3}=1}^{\infty}\frac{n_{2}^{2}}{n_{1}^{\frac{3}{2}}}
(\frac{4\pi^{2}}{\beta^{2}}n_{2}^{2}+\frac{(2\pi)^{2}}{(L_{0}l^{\delta})^{\frac{2}{1+\delta}}}
n_{3}^{\frac{2}{1+\delta}})^{-\frac{1}{4}}\nonumber\hspace{3.5cm}\\
\times
K_{\frac{3}{2}}(2n_{1}R\sqrt{\frac{4\pi^{2}}{\beta^{2}}n_{2}^{2}+\frac{(2\pi)^{2}}{(L_{0}l^{\delta})^{\frac{2}{1+\delta}}}
n_{3}^{\frac{2}{1+\delta}}})\nonumber\hspace{3cm}\\
+\frac{8\sqrt{\pi}}{R^{\frac{1}{2}}\beta^{3}}
\sum_{n_{1},n_{2},n_{3}=1}^{\infty}\frac{n_{2}^{2}}{n_{1}^{\frac{1}{2}}}
(\frac{4\pi^{2}}{\beta^{2}}n_{2}^{2}+\frac{(2\pi)^{2}}{(L_{0}l^{\delta})^{\frac{2}{1+\delta}}}
n_{3}^{\frac{2}{1+\delta}})^{\frac{1}{4}}\nonumber\hspace{4cm}\\
\times[K_{\frac{1}{2}}(2n_{1}R\sqrt{\frac{4\pi^{2}}{\beta^{2}}n_{2}^{2}+\frac{(2\pi)^{2}}{(L_{0}l^{\delta})^{\frac{2}{1+\delta}}}
n_{3}^{\frac{2}{1+\delta}}})\nonumber\hspace{3cm}\\
+K_{\frac{5}{2}}(2n_{1}R\sqrt{\frac{4\pi^{2}}{\beta^{2}}n_{2}^{2}+\frac{(2\pi)^{2}}{(L_{0}l^{\delta})^{\frac{2}{1+\delta}}}
n_{3}^{\frac{2}{1+\delta}}})]\nonumber\hspace{2cm}\\
-4\sqrt{\pi}\frac{R^{\frac{1}{2}}}{\beta^{3}}
\sum_{n_{1},n_{2},n_{3}=1}^{\infty}n_{1}^{\frac{1}{2}}n_{2}^{2}
(\frac{4\pi^{2}}{\beta^{2}}n_{2}^{2}+\frac{(2\pi)^{2}}{(L_{0}l^{\delta})^{\frac{2}{1+\delta}}}
n_{3}^{\frac{2}{1+\delta}})^{\frac{3}{4}}\nonumber\hspace{3.5cm}\\
\times[K_{\frac{1}{2}}(2n_{1}R\sqrt{\frac{4\pi^{2}}{\beta^{2}}n_{2}^{2}+\frac{(2\pi)^{2}}{(L_{0}l^{\delta})^{\frac{2}{1+\delta}}}
n_{3}^{\frac{2}{1+\delta}}})\nonumber\hspace{3cm}\\
+2K_{\frac{3}{2}}(2n_{1}R\sqrt{\frac{4\pi^{2}}{\beta^{2}}n_{2}^{2}+\frac{(2\pi)^{2}}{(L_{0}l^{\delta})^{\frac{2}{1+\delta}}}
n_{3}^{\frac{2}{1+\delta}}})\nonumber\hspace{2cm}\\
+K_{\frac{7}{2}}(2n_{1}R\sqrt{\frac{4\pi^{2}}{\beta^{2}}n_{2}^{2}+\frac{(2\pi)^{2}}{(L_{0}l^{\delta})^{\frac{2}{1+\delta}}}
n_{3}^{\frac{2}{1+\delta}}})]\hspace{2cm}
\end{eqnarray}

\noindent It is obvious that the first term expressed as
$f_{0}=-\frac{\pi^{2}}{240}\frac{1}{R^{4}}$ in Eq.(15) is the same
as the original Casimir pressure on the parallel plates involving
massless scalar fields obeying the Dirichlet conditions without
thermal influence in the four-dimensional spacetime. It should be
pointed out that the correction function represents the deviation
from the fractal additional dimension while the involves thermal
corrections. When the plates are moved far enough away from each
other, the correction function approaches to vanish,

\begin{equation}
\lim_{R\longrightarrow\infty}C(R, \beta,\delta)=0
\end{equation}

\noindent no matter how large or how high the scale dimension or
the temperature is. We have shown that the larger scale dimension
will lead the greater Casimir force [57]. For a definite value of
scale dimension $\delta$ the behaviour of the Casimir pressure on
the variable
$\mu=\frac{R}{(L_{0}l^{\delta})^{\frac{1}{1+\delta}}}$ with
various temperature is plotted in Fig.2. We discover that the
shapes of Casimir force due to different temperature are similar.
The higher temperature lets the curve of the attractive force to
be lower, meaning that the stronger thermal influence will lead
the greater Casimir force between two parallel plates, and the
nature of the Casimir force remains negative. It is interesting
that the Casimir force between the two parallel plates is revised
by both the scale dimension and the surrounding temperature.

\vspace{0.8cm} \noindent \textbf{IV.\hspace{0.4cm}Conclusion}

In this work the Casimir effect for parallel plates at finite
temperature in the spacetime with one fractal extra compactified
dimension is studied. We obtain the total vacuum energy density
according to the Kaluza-Klein model with one additional
non-integer dimension [8]. We regularize the energy density by
means of the regularization of multiple zeta function with one
arbitrary exponent to obtain the Casimir energy density and
further the renormalized Casimir energy density which is the
difference between the Casimir energy densities for two parallel
plates and without plates respectively. It is declared that when
the temperature is high enough the sign of the renormalized
Casimir energy of the system consisting of two parallel plates
will keep negative no matter what the value of scale dimension
$\delta$ is equal to within its own region. In addition to the
fractal extra spatial dimension, the Casimir force has something
to do with the thermal influence. The hotter environment leads the
Casimir force greater. The shapes of Casimir force between two
parallel plates due to different temperature are similar. The two
plates keep attracting each other no matter how high the
temperature is. Our research can be generalized to the case that
the world has more than one multiple fractal additional
compactified spatial dimensions.

\vspace{3cm}

\noindent\textbf{Acknowledgement}

This work is supported by NSFC No. 10875043 and is partly
supported by the Shanghai Research Foundation No. 07dz22020.

\newpage
\begin{figure}
\setlength{\belowcaptionskip}{10pt} \centering
  \includegraphics[width=15cm]{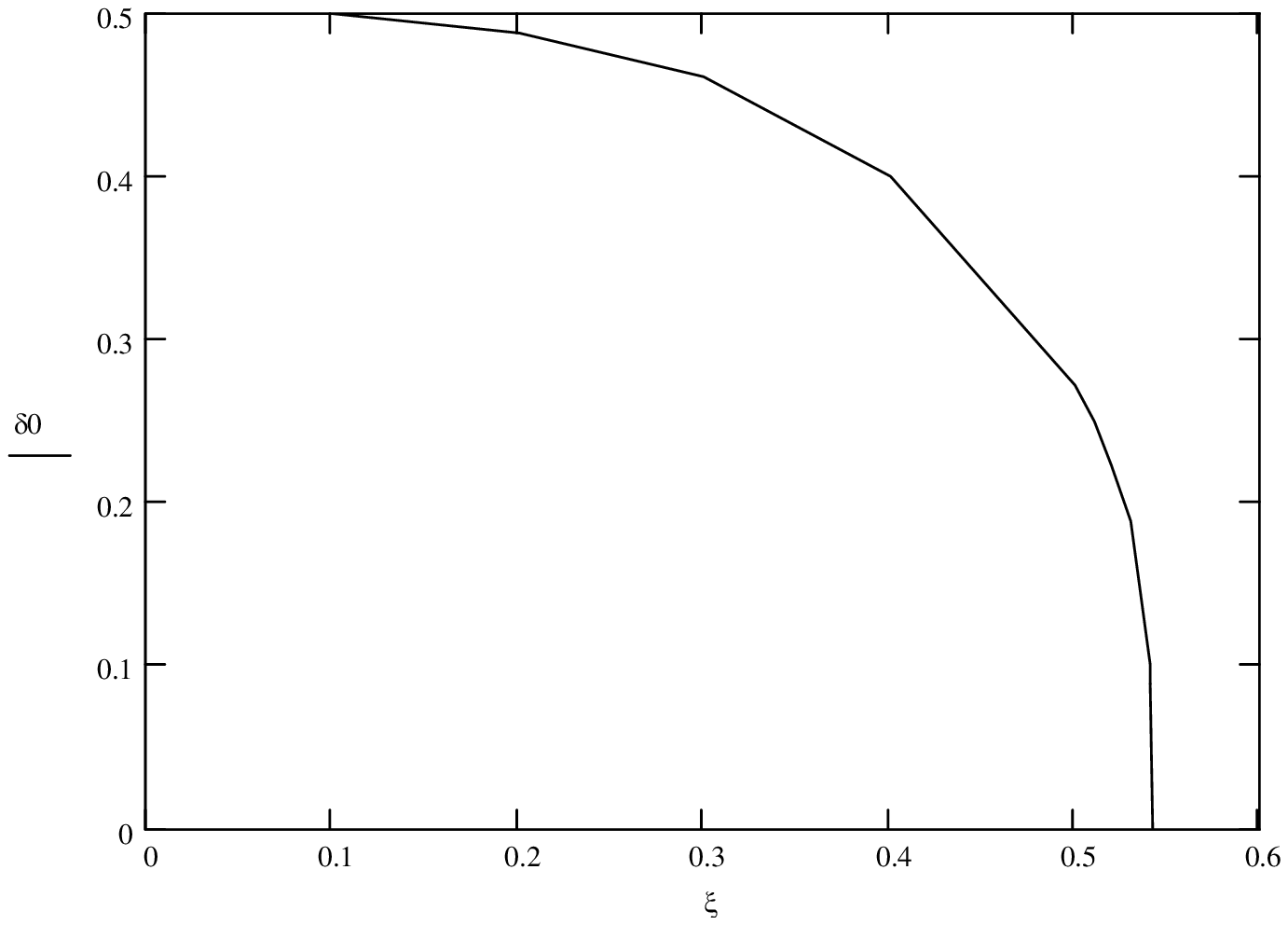}
  \caption{The relation between temperature-dependent ratio
  $\xi=\frac{(L_{0}l^{\delta})^{\frac{1}{1+\delta}}}{\beta}$
  and parameter $\delta_{0}$ and when the scale dimension for extra
  space satisfies $\delta>\delta_{0}$, the sign of Casimir energy for
  parallel plates keeps negative.}
\end{figure}

\newpage
\begin{figure}
\setlength{\belowcaptionskip}{10pt} \centering
  \includegraphics[width=15cm]{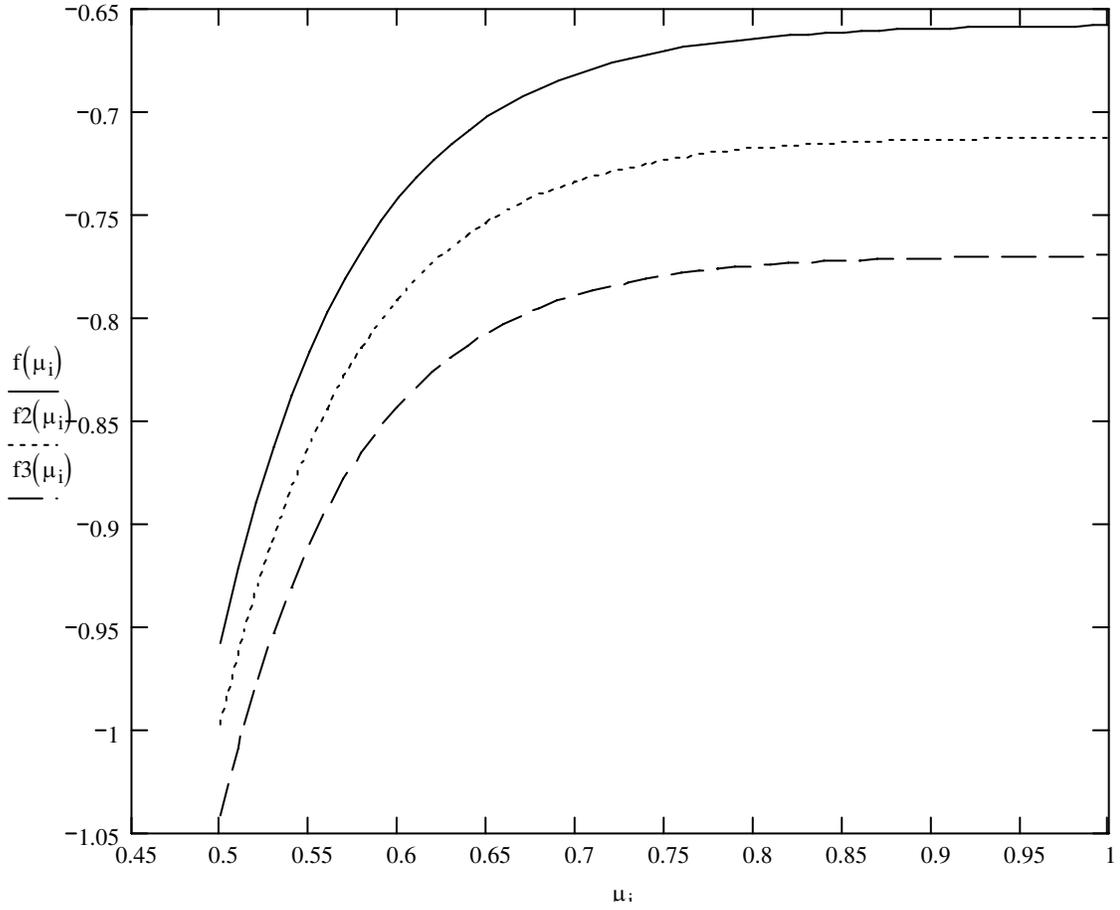}
  \caption{The solid, dot, dashed curves of Casimir force per unit area between two parallel plates
  as functions of ratio $\mu=\frac{R}{(L_{0}l^{\delta})^{\frac{1}{1+\delta}}}$
  at finite temperature with $\frac{(L_{0}l^{\delta})^{\frac{1}{1+\delta}}}{\beta}=1,1.02, 1.04$
  respectively in the presence of one fractal extra compactified dimension for
  $\delta=0.6$.}
\end{figure}

\end{document}